\documentclass[prd,preprint,eqsecnum,nofootinbib,amsmath,amssymb,
               tightenlines,dvips]{revtex4}

\usepackage{color}
\usepackage{epsf}
\usepackage{bm}

\begin{document}

\def\ie{{\rm i.e.}}
\def\red{\color{red}}
\def\Mrowczynski{Mr\'owczy\'nski}
\def\putbox#1#2{\epsfxsize=#1\textwidth \epsfbox{#2}}
\def\centerbox#1#2{\centerline{\epsfxsize=#1\textwidth \epsfbox{#2}}}
\def\half{\frac{1}{2}}
\def\Eq#1{Eq.~(\ref{#1})}
\def\J{{\bm J}}
\def\F{{\bm F}}
\def\E{{\bm E}}
\def\B{{\bm B}}
\def\k{{\bm k}}
\def\p{{\bm p}}
\def\q{{\bm q}}
\def\v{{\bm v}}
\def\be{\begin{equation}}
\def\ee{\end{equation}}
\def\bea{\begin{eqnarray}}
\def\eea{\end{eqnarray}}
\def\alphas{\alpha_{\rm s}}
\def\st{\be}
\def\nd{\ee}
\def\Qs{Q_{s}}
\def\dd{\textrm{d}}
\def\epsilonbath{\epsilon_{\rm bath}}
\def\pmin{p_{\rm min}}
\def\pmax{p_{\rm max}}
\def\nhard{n_{\rm hard}}
\def\favg{f_{\mbox{\scriptsize{angle-avg}}}}
\def\bjork#1{{\left.{#1}\right|_{\rm Bjorken}}}
\def\dyn#1{{\left.{#1}\right|_{\rm dyn}}}
\def\qhat{{\hat{q}}}
\def\qhatinst{\qhat_{\rm inst}}
\def\qhatelast{\qhat_{\rm elastic}}
\def\qhatTelast{\qhat_{T {\rm elast}}}
\def\qhatTinst{\qhat_{T{\rm inst}}}
\def\qhatnew{\qhat_{\mbox{\scriptsize{new-inst}}}}
\def\Gsplit{\Gamma_{\rm split}}
\def\Gmerge{\Gamma_{\rm merge}}
\def\Gammael{\Gamma_{\rm el}}
\def\Gsplitjoin{\Gamma_{\mbox{\scriptsize{split/join}}}}
\def\tbroaden{t_{\rm broaden}}
\def\kfilled{k_{\rm filled}}
\def\kinst{k_{\rm inst}}
\def\kiso{k_{\rm iso}}
\def\ksplit{k_{\rm split}}
\def\kisosplit{k_{\rm isosplit}}
\def\kold{k_{\rm old}}
\def\knewinst{k_{\mbox{\scriptsize{new-inst}}}}
\def\tcoh{t_{\rm coh}}
\def\tlargeang{t_{\rm large\:angle}}
\def\teq{t_{\rm eq}}
\def\tsplit{t_{\rm split}}
\def\tmerge{t_{\rm merge}}
\def\tsplitjoin{t_{\rm split/join}}
\def\tform{t_{\rm form}}
\def\tiso{t_{\rm iso}}
\def\tisofill{t_{\rm iso}(\pmax)}
\def\tisosplit{t_{\rm isosplit}}
\def\tnewinst{t_{\mbox{\scriptsize{new-inst}}}}
\def\tTinst{t_{T_*{\rm inst}}}
\def\tTelast{t_{T_*{\rm elast}}}
\def\tchange{t_{\rm change}}
\def\Tfinal{T_{\rm final}}
\def\att2{\textrm{att2}}
\def\OO{{\cal O}}
\def\gsim{\mbox{~{\raisebox{0.4ex}{$>$}}\hspace{-1.1em}
        {\raisebox{-0.6ex}{$\sim$}}~}}
\def\lsim{\mbox{~{\raisebox{0.4ex}{$<$}}\hspace{-1.1em}
        {\raisebox{-0.6ex}{$\sim$}}~}}

\title{Bjorken Flow, Plasma Instabilities, and Thermalization}

\author{Aleksi Kurkela and Guy D.\ Moore}
\affiliation
    {%
    Department of Physics,
    McGill University, 3600 rue University,
    Montr\'eal QC H3A 2T8, Canada
    }%

\date{August 2011}

\begin{abstract}
    {%
      At asymptotically high energies, thermalization in heavy ion
      collisions can be described via weak-coupling QCD.  We present a
      complete treatment of how thermalization proceeds, at the parametric
      weak-coupling level.  We show that
      plasma instabilities dominate the dynamics, from immediately after
      the collision until well after the plasma becomes nearly in
      equilibrium.  Initially they drive the system close to isotropy,
      but Bjorken expansion and increasing diluteness makes the system
      again become more anisotropic.
      At time $\tau \sim \alpha^{\frac{-12}{5}} \Qs^{-1}$ the dynamics
      become dominated by a nearly-thermal bath; and at
      time $\tau \sim \alpha^{\frac{-5}{2}} \Qs^{-1}$ the bath comes to
      dominate the energy density, completing thermalization.  After
      this time there is a nearly isotropic and thermal Quark-Gluon
      Plasma.
    }%
\end{abstract}

\maketitle
\thispagestyle {empty}

\section{Introduction and Summary}

Relativistic heavy ion experiments have given us new insights into the
dynamics of hot QCD \cite{expt1,expt2}.  One of the most striking
results is the success of a hydrodynamic description
\cite{oldhydro,newhydro}.
This seems to imply that the hot QCD matter generated in a heavy ion
collision thermalizes quite quickly and then has a period as a nearly
thermal medium well described by hydrodynamics (which is predicated on
the assumption of local near-equilibrium).  Theoretically, we have no
idea why this should be true.  In fact, we do not even understand early
thermalization in the theoretically clean case where we consider the
limit of large nuclei at arbitrarily high energy per nucleon.  In this
limit, the very early dynamics are believed to be well understood, in
the framework of the ``colored glass condensate''
\cite{coloredglass,coloredglass2} and its post-collision debris, the ``glasma''
\cite{glasma}.  According to this description, the initial conditions
should be approximately boost-invariant, intense classical colored fields
with a single energy scale $\Qs$, the ``saturation scale''.  Boost
invariance means that the natural coordinates are the transverse
directions $x,y$, rapidity $\eta$, and proper time since the collision,
$\tau$.  We will rename $\tau$ to $t$ and generally work in terms of
$z \equiv \eta \tau$, the physical distance along the beam direction in
$\eta,\tau$ coordinates,
which expands (red-shifts) linearly with time (Bjorken expansion).  The
scale $\Qs$ increases with increasing nuclear size and with increasing
energy per nucleon \cite{smallx}, so in principle this scale becomes
large in the limit of interest (though whether this is true in practice
at achievable energies is another matter).  Provided that
$\Qs \gg \Lambda_{\rm QCD}$, physics at the scale $\Qs$ should be characterized
by weakly-coupled QCD, with the coupling taking a characteristic value
$\alphas(\Qs) \ll 1$ (henceforth we write $\alphas(\Qs)$ as $\alpha$).  Weak
coupling should make the theoretical problem easier to address.
Nevertheless, the subsequent evolution of these fields is not well
understood.

The most comprehensive attempt to date to describe thermalization
at central rapidities
from these initial conditions was by Baier, Mueller, Schiff and Son (BMSS)
\cite{bottomup}.  They argued that the medium made up of $p\sim \Qs$
excitations becomes anisotropic and dilute; but LPM suppressed radiation
of lower energy daughters generates a bath of $p \ll \Qs$ excitations which
come to dominate the dynamics and lead to the breakup of the $p\sim \Qs$
``hard'' ``parent'' excitations into a thermal bath.  However, as pointed
out by Arnold, Lenaghan and Moore \cite{ALM}, their discussion ignored
the physics of plasma instabilities \cite{Mrow,RS},
which should dominate at least the
early stages of the dynamics.  The role of plasma instabilities in these
early stages was considered analytically by B{\"o}deker \cite{Bodeker}
and by a mix of analytic and numerical considerations by Arnold and Moore
\cite{ArnoldMoore2007}.  Classical field simulations with plasma
instabilities were considered by Romatschke and Venugopalan
\cite{RV}, and several groups made related studies
\cite{several}.  However, none of these studies followed the dynamics
past the early stages or explained how final thermalization occurs.

In this paper we will take advantage of our recent study of the dynamics
of weakly coupled gauge theories out of equilibrium (Ref.~\cite{KM1},
henceforth ``KM1'') to give a purely analytical and parametric description
of how ``the glasma'' should thermalize when $\Qs$ is such a large scale
that $\alpha \ll 1$.  The initial state of the glasma should be well
described by intense classical fields with much longer coherence in the
$\eta$ direction than in the transverse $xy$ plane \cite{glasma}.  That
is, the initial conditions are very anisotropic, with a longer coherence
length (smaller mean wave number) in the $\eta$ direction.  But we find
(as suggested by Ref.~\cite{RV}) that plasma instabilities rapidly
reorganize this into a configuration which is only anisotropic at the
$\OO(1)$ level.

From this point we find {\sl two} attractor solutions.  
In one solution,
the plasma becomes highly anisotropic.
Bjorken expansion increases the anisotropy but does not
lower the wave-number of typical excitations.  This leads to a
dilute plasma of wave-number $p\sim \Qs$ excitations.  This
solution is similar to the one found by BMSS
\cite{bottomup}, except that plasma instabilities dominate the dynamics
(as proposed in Ref.~\cite{ALM} and discussed in
Refs.~\cite{Bodeker,ArnoldMoore2007}).  In this solution the momentum
space anisotropy grows with time as $t^{\frac 18}$.
In the other attractor solution, 
the plasma becomes nearly isotropic at a time $t$ longer than $\Qs^{-1}$
by at most logarithms of $\alpha$.  Plasma instabilities resist Bjorken
expansion to keep the plasma nearly isotropic.  The typical excitation
energy falls with time, and the typical occupancy falls but not as fast
as in the first solution.
This solution is somewhat like the one recently suggested
by Blaizot, Gelis, Liao, McLerran, and Venugopalan \cite{BGLMV}, except
that we find that plasma instabilities play a key role in the dynamics
and that no Bose-Einstein condensate forms.  
Fully nonperturbative and nonequilibrium early-time dynamics
determine which attractor the system finds.  But we argue that the path
from the initial conditions to the more anisotropic attractor is much
more likely.

Assuming the first (strongly anisotropic) attractor is correct, we find
that a nearly-thermal bath of low-momentum ($p\sim T \ll \Qs$)
excitations forms and comes to dominate the dynamics at time
$t\sim \alpha^{\frac{-12}{5}} \Qs^{-1}$.  By 
$t\sim \alpha^{\frac{-5}{2}} \Qs^{-1}$ the soft bath causes the ``hard''
$p \sim \Qs$ excitations to disintegrate and the system becomes nearly
isotropic and thermal, with a temperature
$\Tfinal \sim \alpha^{\frac{-11}{24}} \Qs^{\frac{2}{3}} t^{\frac{-1}{3}}$.
But while nearly thermal, the system is initially non-Newtonian;
anomalous viscosity \cite{anom-visc} (due to plasma instabilities)
controls the level of isotropy until
$t\sim \alpha^{\frac{-45}{16}} \Qs^{-1}$, after which the fluid becomes
Newtonian.  Before $t \sim \alpha^\frac{-5}{2} \Qs^{-1}$ the plasma is
far from equilibrium; we parametrically estimate the (quite
complicated) phase space particle number density at all momenta $p \leq
\Qs$ and all times from $\Qs^{-1}$ to final thermalization.  The results
are presented in Figure~\ref{fig1} and Table \ref{table1}.

Our results indicate a faster
breakup of the hard $p\sim \Qs$ excitations than the old estimate
$t\sim \alpha^{\frac{-13}{5}} \Qs^{-1}$
from BMSS \cite{bottomup}, but still somewhat slower than the absolute
lower bound of $\alpha^{\frac{-7}{3}} \Qs^{-1}$ derived by Arnold and
Lenaghan \cite{AL}. But it is consistent with the estimate for full
thermalization times presented in KM1, who claim
$\teq \sim \alpha^{\frac{-13}{7}} \Qs^{\frac 57} T^{\frac{-12}{7}}$;
inserting $\Tfinal$ returns $t\sim \alpha^{\frac{-5}{2}} \Qs^{-1}$.

\section{Initial Conditions and Attractors}

As we briefly mentioned in the introduction, an extensive literature on
the small-$x$ behavior of large nuclei (see Ref.~\cite{smallx}) and
small-$x$ evolution in heavy ion collisions \cite{coloredglass,glasma}
indicates that, at time $t \sim \Qs$, the debris of a heavy ion collision
is dominated by classical gluon fields with coherence length
$l\sim 1/\Qs$ in the $xy$ plane, much longer coherence in $z=\eta t$,
and energy density
$\varepsilon(t\sim \Qs^{-1}) \sim \alpha^{-1} \Qs^{4}$. At times
$t > \Qs^{-1}$ these fields have lost phase coherence and the evolution
should be described with the tools we introduced in KM1, modified to
include the Bjorken expansion of the geometry.

\subsection{Combining KM1 with Bjorken expansion}

\begin{table}
\centerline{\begin{tabular}{|c|c|c|} \hline
Variable & Describes & Detailed Definition \\ \hline
$a$ & time & $t = \alpha^{-a} \Qs^{-1}$ \\ \hline
$b$ & Typical momentum & $Q \sim \alpha^b \Qs$, where 
        $f(p\gg Q) \ll f(p\sim Q)$ \\ \hline
$c$ & Mean occupancy & $f(\p=Q \hat{x}) \sim \alpha^{-c}$ \\ \hline
$d$ & Anisotropy & 
$\begin{array}{ccl}
d<0: & & f(Q\hat{x}) - f(Q\hat{z})
                          \sim \alpha^{-d} f(Q) \\
 d>0: & & f(p_z,\p_\perp) \ll f(0,\p_\perp)
     \; \mbox{ if }\;  p_z > \alpha^{d} |\p_\perp| \\ \end{array}$
 \\ \hline
\end{tabular}}
\caption[Our irritating notation]
{\label{our_notation} Brief summary of our log notation for the most
important descriptors of the system}
\end{table}

We begin by introducing notation, including the notation used in KM1.
In KM1 we considered the evolution of a system with a single dominant
characteristic wave number $Q$ in a time-independent geometry.
The system is characterized by a scale $Q$ (which need
not be $\Qs$; we will write $Q = \alpha^{b} \Qs$), a typical occupancy
$f(p\sim Q) \sim \alpha^{-c}$, and a degree of anisotropy characterized
by $d$.  For $d>0$ most excitations have $p_z \lsim \alpha^{d} p$
and we write $\delta \equiv \alpha^d$.  The occupancy 
$f(p)\sim \alpha^{-c}$ applies for $p_z < \delta p$, outside this range
the occupancy is small, so the particle number and energy densities are
\be
\label{n_d>0}
\mbox{for $d>0$, }\quad
n \sim \alpha^{d-c} Q^3 \;\; \mbox{and} \;\;
\varepsilon \sim \alpha^{d-c} Q^4 \sim \alpha^{d-c+4b} \Qs^4 \,.
\ee
For $d<0$ the particle number is nearly
direction independent, but the direction dependence varies with relative
amplitude $\epsilon \equiv \alpha^{-d}$ -- that is,
$( f(Q \hat{x}) - f(Q \hat{z}) )/f(Q) \sim \epsilon$.  Therefore,
for $d<0$, the particle number and energy densities are
\be
\label{n_d<0}
\mbox{for $d<0$, }\quad
n \sim \alpha^{-c} Q^3 \;\; \mbox{and} \;\;
\varepsilon \sim \alpha^{-c} Q^4 \sim \alpha^{-c+4b} \Qs^4 \,.
\ee
We will also introduce one more logarithmic variable; we will
write the time as $t = \alpha^{-a} \Qs^{-1}$ (or 
$a \equiv \ln(t\Qs)/\ln(1/\alpha)$).  The initial conditions
are fixed at $t\sim \Qs^{-1}$, which is $a=0$.  Parametrically later
times are $a>0$.
Our notations are summarized in Table \ref{our_notation}.

\begin{figure}
\centerbox{0.8}{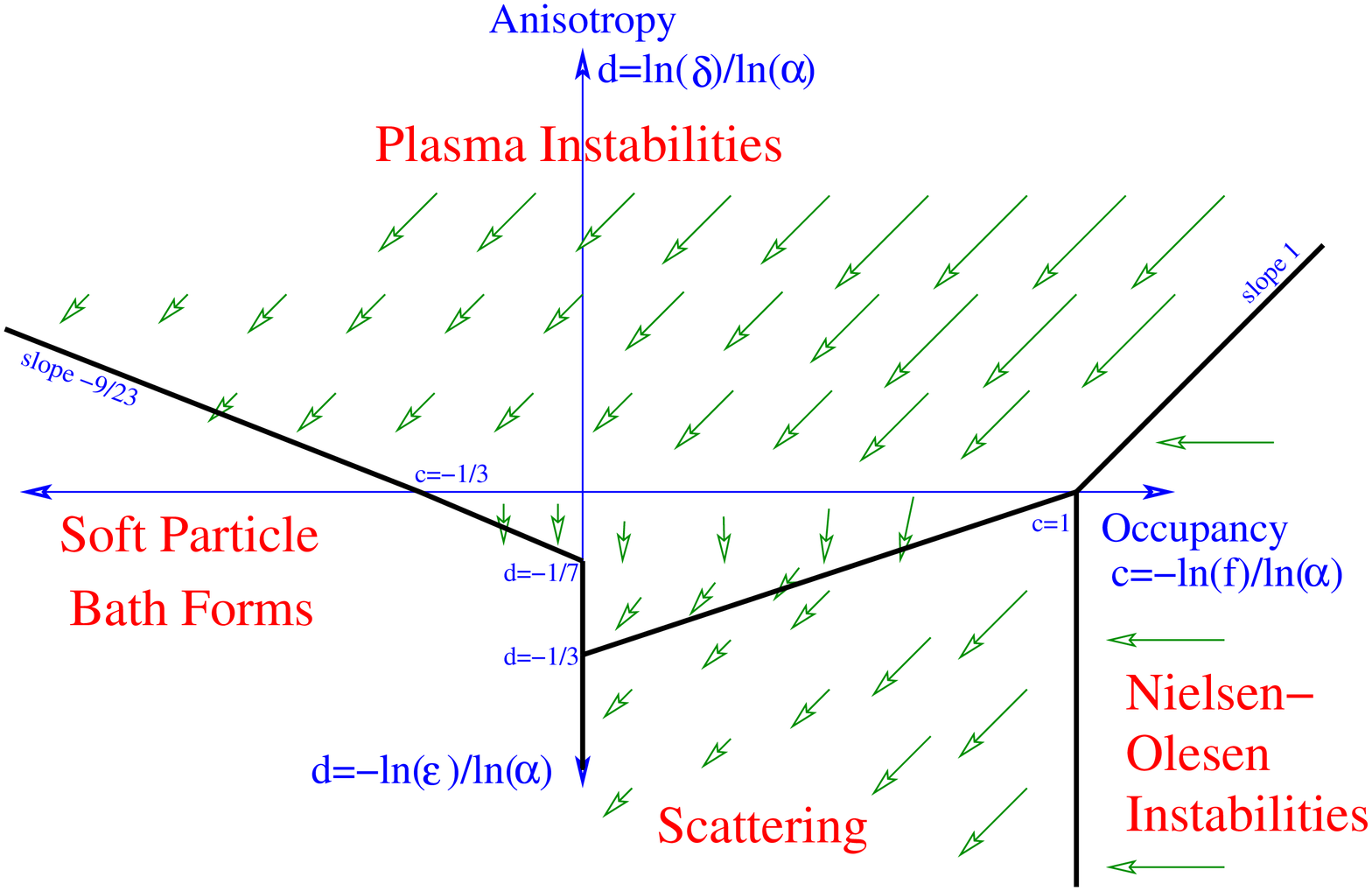}
\caption[Simplified figure from Paper 1]
{\label{KM1fig1}
  Simplified version of Figure 1 from KM1 \cite{KM1}.  The
  vertical axis represents how anisotropic the dominant excitations
  are; the horizontal axis shows how large the dominant excitation
  occupancies are.  The labels indicate what the dominant physics is in
  each region; the arrows suggest how those dynamics ``move'' the system
  in this two-dimensional space.  There are no arrows in the ``soft
  particle bath'' region because here the typical excitations break up
  and are replaced by something very different.}
\end{figure}

According to KM1, the occupancy--anisotropy or $c,d$ plane is divided into
four regions, shown in Figure \ref{KM1fig1}.  At large anisotropy the
dynamics are dominated by magnetic or ``Weibel'' \cite{Weibel}
plasma instabilities.
That means that certain long-wavelength magnetic fields grow to
large amplitudes; there are associated electric fields but they are much
weaker.  Magnetic fields change an excitation's direction without
changing the energy; so the main physics in this region involves the
randomization of excitations' directions.  Induced (bremsstrahlung)
emission or merging of excitations also occurs but is subdominant.
For less isotropic
but over-occupied systems ($1>c>0$ and $d<(c-1)/3$), ordinary scattering
dominates the dynamics.  Elastic scattering drives the system towards
isotropy, and inelastic scattering drives the mean occupancy $c$ towards
zero.  The two types of scattering are of comparable efficiency
(in stark contrast to scalar field theory).  For underoccupied systems
below some level of anisotropy, the main physics is that a bath of
smaller-momentum excitations forms and eventually causes the dominant
excitations to shower and join this bath.  The last region, of extremely
high occupancies, exhibits Nielsen-Olesen instabilities \cite{NO}.  This
region has no relevance for us.  The arrows in Figure \ref{KM1fig1} show
qualitatively how these dynamics cause $c,d$ to evolve with time.

All this is for a homogeneous and non-expanding system.  But the debris
of a heavy ion collision expand.  In our choice of coordinates the
system is nearly statistically homogeneous, but the $z$ direction grows
linearly with time.  This Bjorken expansion red-shifts away the
$z$ component of momentum, increasing the anisotropy of the system, and
raising $d$.  And for $d<0$ it lowers the typical excitation's energy,
increasing $b$.  In
the scattering, plasma unstable, and Nielsen-Olesen regions the evolution
of $b,c,d$ will be determined by an interplay of the equilibration
processes and Bjorken expansion.

\subsection{Initial conditions and attractors}

\begin{figure}
\centerbox{0.75}{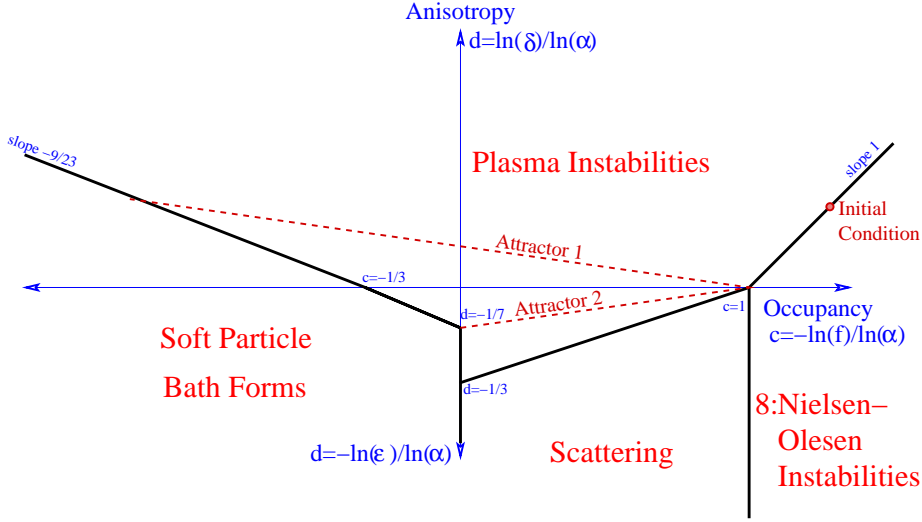}
\caption{\label{attractors}
Occupancy-anisotropy plane, the initial conditions from heavy ion
collisions, and the two attractor solutions we find.}
\end{figure}

The initial conditions we described above have $d>0$ and $c=1+d$
at time $t \sim \Qs^{-1}$.  This is the point marked ``Initial
Condition'' in Figure~\ref{attractors}.  The figure also shows two
attractor solutions which we find, which we now explain in more detail.
By an attractor we mean a path on the $c,d$ plane, such that a system
relatively close to the path will fall onto and then follow the path as
a result of the combination of dynamics and Bjorken expansion.  There
are two attractors because there is a fundamental difference between the
dynamics of a system with large anisotropy versus a system with small
anisotropy.  This difference turns out to allow solutions of either
type.

Consider first a system which is highly anisotropic, so the typical
excitation has $p_z \ll |\p|$.  Then Bjorken expansion does not change
the typical excitation's energy.  And angle randomization -- such as
that induced by plasma instabilities -- increases the region of phase
space which is occupied, and therefore brings down the typical
occupancy (reduces $c$).  But consider instead a system which is only weakly
anisotropic.  Typical excitations have $|p_z| \sim |\p|$, and therefore
Bjorken expansion reduces typical excitation energy.  And since angle
randomization is mixing modes of nearly the same occupancy, it has
little impact on the typical occupancy.  These fundamental differences
mean that we have to treat the two cases, $d>0$ and $d<0$, separately.
And we will find an attractor in each.  Specifically, we will find
for $d>0$,
\be
d = \frac{a}{8} \,, \quad
c = 1-\frac{7a}{8} \,, \quad
b = 0 \,, \qquad \mbox{(attractor 1)}
\label{attract1}
\ee
while for $d<0$ there is another attractive solution,
\be
d = -\frac{8a}{135} \,, \quad
c = 1-\frac{56a}{135} \,, \quad
b = \frac{31a}{135} \,. \qquad \mbox{(attractor 2)}
\label{attract2}
\ee
These attractors, and the initial conditions, are displayed in
Figure \ref{attractors}.

\subsection{First (anisotropic) attractor}

First consider the case $d>0$, where most excitations have $p_z \ll
|\p|$.  This region lies entirely in the plasma
instabilities dominated region in Figure \ref{KM1fig1} provided $c>-1/3$
(as it initially is).  To describe evolution in this region it is
convenient to think of the descriptors of a system, $b$, $c$, and $d$,
as functions of time or of $a$, and to ask about their $a$ derivatives;
$c' \equiv \dd c(a)/\dd a$ and similarly for $b'$, $d'$.
Each quantity evolves due to Bjorken expansion and due to the effects of
dynamics, principally plasma instabilities; we will write
$c' = \bjork{c'} + \dyn{c'}$ to distinguish the part of the evolution
from each effect.

The effect of Bjorken expansion is a red-shifting of the $p_z$ component
of momentum,
\be
\frac{\dd p_z}{\dd t} = -\frac{p_z}{t} \qquad \mbox{(Bjorken expansion)}\,.
\label{eq:bjorken}
\ee
This does not change $c$; occupied states retain their occupancy, they
just change the value of momentum.  It also does not change $b$; since
$p_z \ll |\p|$, the typical excitation's energy is not changed.
Thus, $\bjork{c'} = 0 = \bjork{b'}$.
But Bjorken expansion does lead to an increase in $d$.  Recall that
$\delta \equiv (p_z/p)_{_{\rm RMS}} = \alpha^d$.  The angle $\delta$
evolves according to
\be
\frac{\dd\delta}{\dd t} = \frac{\dd\delta}{\dd p_z} \frac{\dd p_z}{\dd t}
= \frac{\delta}{p_z} \frac{\dd p_z}{\dd t} = -\frac{\delta}{t} \,,
\label{delta_ev}
\ee
which using \Eq{eq:bjorken} and the definitions of $a,d$ becomes
\be
\label{bjorkd1}
\bjork{d'} = \frac{\dd (d)}{\dd a} = \frac{\dd (d)}{\dd t} \frac{\dd t}{\dd a}
= \frac{\dd (\ln(\delta)/\ln(\alpha))}{\dd t}
  \frac{\dd t}{\dd (\ln(t\Qs)/\ln(\alpha^{-1}))}
= -\frac{t \dd \delta}{\delta \dd t} = +1 \,.
\ee

Now we need to add the effect of plasma instabilities.  As mentioned,
the dominant thing the instabilities do is to deflect excitations,
widening their angular range.  This changes $c$ and $d$, but not the
typical energy $b$; $\dyn{b'}=0$.  Since $\bjork{b'}=0$ as well, we find
$b'=0$ and therefore $b=0$; the typical excitation energy remains
$Q=\Qs$.
Dynamics also cannot change
the total energy density $\varepsilon$.
According to \Eq{n_d>0}, $\varepsilon$ depends on the combination
$d-c+4b$.  Since $b'=0$, we learn that
$\dyn{d'} - \dyn{c'} = 0$.  Combining with \Eq{bjorkd1}, we learn
that $d'-c' = 1$.  Since initially $c-d=1$, we have
\be
b=0\,, \quad c-d = 1-a \qquad \mbox{($d>0$ case)} \,.
\label{c_d>0}
\ee
Therefore the state of the system at a given time $t$ (or a given value
of $a$) is determined by $d(a)$; $c$ is fixed to be $c=1+d-a$.

To finish determining the dynamics we need to find out how fast plasma
instabilities cause the level of anisotropy to change.  According to
KM1, there is a time scale $\tbroaden$ which determines how long it
takes the plasma instabilities to significantly increase the range of
angles of $p\sim Q$ excitations (to reduce $d$).  The anisotropy
therefore evolves at
\be
\frac{\dd (d)}{\dd t} \sim -\frac{1}{\tbroaden} \qquad \Rightarrow \qquad
\dyn{d'} = \frac{-t}{\tbroaden} = -\alpha^{-a} \Qs^{-1} \tbroaden^{-1} \,.
\label{dprime1}
\ee
According to KM1 \cite{KM1}, the time scale $\tbroaden$ is
\be
\tbroaden \sim \left\{ \begin{array}{ll}
\alpha^{\frac{c-d-1}{2}} Q^{-1} & d > \frac{1-c}{3}\,, \\
\alpha^{\frac{3c+5d-3}{2}} Q^{-1} & d < \frac{1-c}{3}\,. \\
\end{array} \right.
\label{dprime2}
\ee
Combining with \Eq{c_d>0} and \Eq{dprime1}, we find
\be
\dyn{d'} \sim \left\{ \begin{array}{ll}
-\alpha^{\frac{-a}{2}} & d > \frac{1-c}{3} \,, \\
-\alpha^{\frac{a-8d}{2}} & d < \frac{1-c}{3} \,. \\
\end{array} \right.
\label{dprime3}
\ee

At a time $a$ during the evolution, $d$ might take its value on, above,
or below the attractor.  Since we have two constraints, \Eq{c_d>0},
on $b,c,d$, the possible deviation from the attractor is completely
fixed by the value $d(a)$.
If $d(a)>a/8$ then $\dyn{d'} \ll -1$ and $d$ falls very rapidly, on a time
scale shorter than the age of the system.  If $d<a/8$ then
$\dyn{d'} \simeq 0$ and $d' = 1$.  In this case $d$ grows linearly with
$a$, rapidly exceeding $a/8$.  Therefore the solution $d=a/8$ is an
attractor.  The rest of \Eq{attract1} follows from \Eq{c_d>0}.
According to Figure~\ref{attractors}, this
solution makes sense until it reaches the point $d=3/10$, $c=-11/10$,
which occurs at $a=12/5$ ($t\sim \alpha^{\frac{-12}{5}} \Qs^{-1}$).
After this time a soft bath forms and takes over the dynamics.  We will
return to these ``late-time'' dynamics in the next section.

\subsection{Second attractor}

Now consider the case $d<0$, that is, systems which are nearly
isotropic.  Now \Eq{eq:bjorken} is still correct, but since
$\langle p_z^2\rangle  \simeq \p^2/3$, expansion now changes the typical
particle energy,
\be
\bjork{b'} = 1/3 \,.
\ee
Correspondingly, the energy density falls with time as
$\varepsilon \sim t^{-4/3}$ ($t^{-1}$ due to dilution and $t^{-1/3}$ due
to the work done by longitudinal pressure).  Then \Eq{n_d<0} becomes
(compare with \Eq{c_d>0})
\be
-c'+4b' = 4/3 \quad \mbox{or} \quad
4b = c - 1 + 4a/3 \,.
\label{c_d<0}
\ee
We can use this expression to eliminate $b$ in favor of $c,d$ in what
follows.

Next we find the effect of expansion on $d$.
A reduction of $p_z$ by $p_z \rightarrow p_z(1-\epsilon)$ would induce
an anisotropy of $\epsilon$.  Since 
$d \equiv \ln(\epsilon)/\ln(1/\alpha)$, $d$ is very sensitive to
expansion:
\be
\bjork{d'} \sim \epsilon^{-1} = \alpha^{d} \gg 1 \,.
\label{dprime_2}
\ee
In any self-consistent solution, $\dyn{d'}$ will have to be
compensatingly large, so $d' = (\bjork{d'}+\dyn{d'}) \lsim 1$.

We can immediately rule out the existence of an attractor solution in
the scattering region, $d<(-1+c)/3$.  In this region, number-changing
processes are as efficient as direction-changing elastic processes.%
\footnote{%
    Since there is some confusion in the literature on this point
    \cite{BGLMV}, we will summarize why this is true.  The rate
    for an excitation to undergo a generic-angle $2\leftrightarrow 2$
    process is $\alpha^{2-2c} Q^{-1}$ ($\alpha^2$ for the two vertices,
    $\alpha^{-2c}$ for two Bose factors).  Small-angle scattering is
    enhanced by a factor $Q^2/m^2 \sim \alpha^{-1+c}$ because of the
    Coulomb divergence; but a small angle scattering does little to
    drive the system towards equilibrium.  Adding a hard radiation or
    merging onto a small-angle scattering process reduces the rate by a
    factor of $\alpha^{1-c}$ ($\alpha$ for the vertex, $\alpha^{-c}$ for
    a Bose stimulation factor for the new external state; phase space
    factors are canceled by collinear factors as usual in ISR or FSR),
    yielding a
    rate of $\alpha^{2-2c} Q^{-1}$.  If $f(p\ll Q) \propto p^{-1}$ then
    the rate in the infrared is actually larger by a factor of
    $Q/p$, which makes the development of a negative chemical potential
    impossible (contrary to the assumptions of \cite{BGLMV}).  This
    point is discussed at some length in Ref.~\cite{ADM}, section IIIE.%
    }
Therefore
$\dyn{c'} \sim \dyn{d'}$.%
\footnote{The only exception is if $c=0$, so number changing processes
  are in equilibrium.  That is, there can be nearly-thermal, scattering
  dominated systems, just not ones with parametrically large occupancy.}
So in the scattering region, either $\dyn{d'} \sim 1$, in which case
$d$ rises very fast due to $\bjork{d'} \gg 1$, or
$\dyn{d'} \gg 1$, in which case $\dyn{c'} \gg 1$ and the system evolves
very fast to lower occupancy.  Since lower occupancy features slower
dynamics, both $\dyn{c'}$ and $\dyn{d'}$ would then collapse, again
allowing $d$ to rise into the plasma-unstable region.

So consider the region with important plasma instabilities, 
$0>d>(-1+c)/3$.  According to KM1, there are two relevant time scales
here:  $\tbroaden$, the time scale for plasma instabilities to cause
large-angle direction change; and $\tmerge$, the time scale for
plasma-instability induced number-changing processes.  (Think of these
as hard synchrotron radiation, induced as the excitations bend in the
magnetic fields; but for over-occupied systems the synchrotron
absorption process is more important than emission.)  The time scales
are (also using \Eq{c_d<0})
\bea
0>d>\frac{c-1}{3} \,: \qquad
\tbroaden & \sim & \alpha^{\frac{3d-3+3c}{2}} Q^{-1}
            \sim \alpha^{\frac{3d-3+3c-2b+2a}{2}} t
            \sim \alpha^{\frac{18d+15(c-1)+8a}{12}} t
\,, \nonumber \\
\tmerge& \sim & \alpha^{\frac{3d-7+7c}{4}} Q^{-1}
          \sim \alpha^{\frac{3d-7+7c-4b+4a}{4}} t
           \sim \alpha^{\frac{9d+18(c-1)+8a}{12}} t \,.
\label{tbroad_weak}
\eea
Merging (two particles joining into one with the sum of the initial
particles' energy) raises the typical particle energy and lowers the
occupancy.  So
\be
0>d>\frac{c-1}{3} \,: \qquad
\dyn{c'} \sim -\frac{t}{\tmerge} \,.
\label{c_change}
\ee

Broadening leads towards isotropy, lowering $d$.  But so does merging;
as discussed in KM1, when merging is induced by plasma instabilities,
the particles near the $xy$ plane merge more often than those with large
$p_z$, reducing the number of in-plane excitations.  This last effect is
independent of the anisotropy $\epsilon$, while angle randomization only
affects the $\OO(\epsilon)$ angle-anisotropic part of the distribution.
Therefore
\cite{KM1}
\be
\dyn{d'} = - \left(
  \frac{t}{\tbroaden} + \alpha^d
  \frac{t}{\tmerge} \right) \,.
\label{d'2}
\ee
The $\tbroaden$ term dominates if $0>d>(-1+c)/7$; for $d<(-1+c)/7$ the
$\tmerge$ term dominates.

\begin{figure}
\centerbox{0.5}{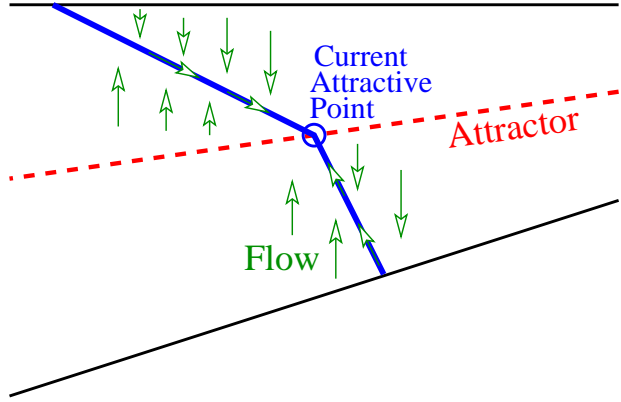}
\caption[Second attractor detail]
{\label{fig_attract2}
 Evolution near the second attractor.  At a given time $a$, one point on
 the attractor line (circled in blue) is preferred.  Above the solid
 blue line, the dynamics rapidly lowers the anisotropy; above the line,
 Bjorken expansion rapidly raises it (as suggested by green arrows).
 On the blue line, number-changing processes move the system towards the
 blue circled point.}
\end{figure}

Now we show that \Eq{attract2} constitutes an attractor.  While in the
$d>0$ case we had two constraints on $b,c,d$, now we have only one
constraint, \Eq{c_d<0}.  So if we are off the attractor at time $a$,
we are at some point in the $c,d$ plane.  Figure
\ref{fig_attract2} shows how $c,d$ might deviate from the attractor
solution, \Eq{attract2}, at some given time $a$.
The blue line is the line where $-\dyn{d'}=\bjork{d'}$:  using
\Eq{dprime_2}, \Eq{d'2} and \Eq{tbroad_weak},
\be
-\dyn{d'}=\bjork{d'} \quad \mbox{if} \quad
\left\{ \begin{array}{ll}
  d=\frac{1-c}{2}-\frac{4a}{15}\,, & d>\frac{c-1}{7}\,, \\
  d=2(1-c)-\frac{8a}{9} & d<\frac{c-1}{7} \,. \\
  \end{array} \right.
\label{blueline}
\ee
Above this line, $|\dyn{d'}| > \bjork{d'}$.  Since both are large, $d'$
is large and negative here.  Below the line, $|\dyn{d'}| < \bjork{d'}$
and $d'$ is large and positive.  In either case the dynamics drive $d$ to
the line on a time scale short compared to the system's age.

Next we must check the evolution of $c$ on the blue line in Figure
\ref{fig_attract2}, to see whether it is driven towards the attractor.
To the left on the line, where $d>d_{\rm attract}$ and
$c<c_{\rm attract}$, \Eq{tbroad_weak}, \Eq{c_change}, and
\Eq{blueline} show that $c'\simeq 0$.  Since the attractor moves towards
lower $c$ with time (\Eq{attract2}), points here move towards being on
the attractor.  To the right of the line, where $c>c_{\rm attract}$,
\Eq{c_change} and \Eq{d'2} show that 
$\dyn{c'} \sim \alpha^d \dyn{d'} \sim -1$.  In fact it is more negative
than -1, since every merging process changes typical occupancy, but only
the anisotropic part of merging processes change anisotropy.  Therefore
$c$ would fall quickly when it is greater than the attractor value.
This verifies that the solution, \Eq{attract2}, constitutes an attractor.

\subsection{Why we think the first attractor is realized}

The initial conditions are $c=1+d$ and $d>0$, corresponding to an energy
density $\varepsilon \sim \alpha^{-1} \Qs^4$.  Using \Eq{dprime3},
as soon as $a>0$ this initial condition should fall rapidly to
approximately the point $c=1,d=0$.  That is the point where the two
attractors meet.  So which attractor will the system actually follow?

\begin{figure}
\centerbox{0.5}{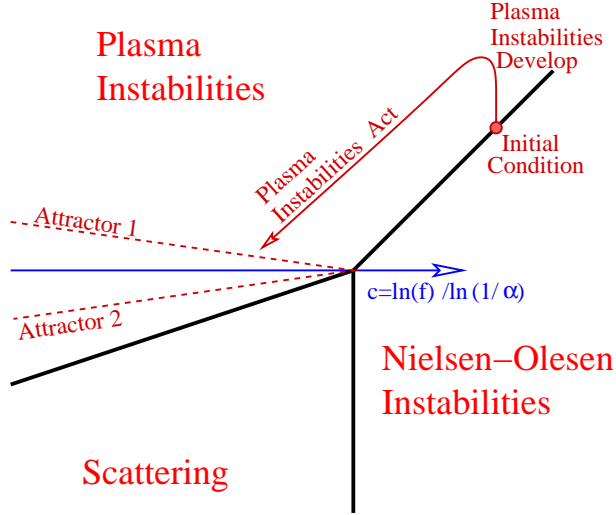}
\caption{\label{fig_logs}
  Zoom-in on the region of the $c,d$ plane near $c=1,d=0$, showing the
  early-time evolution from the initial conditions.  It takes some time
  for the instabilities to develop, during which Bjorken expansion
  increases anisotropy.  When plasma instabilities finally act, they
  bring the system to a region where the two
  attractors are well separated, leading to attractor 1.}
\end{figure}

Here we argue that, if we try to include corrections suppressed by a
logarithm of $\alpha$, it appears that the first (anisotropic, $d>0$)
attractor is
preferred.  Consider first the initial conditions.  The characteristic
wave-number (or inverse coherence length) is $\Qs$, the energy density
is $\varepsilon \sim \alpha^{-1} \Qs^3/t$.
Under these conditions plasma instabilities grow
at a characteristic rate called $m$, the screening scale \cite{RS};
for these conditions $m^2 \sim \alpha \varepsilon / \Qs^2 \sim \Qs/t$.
The unstable modes have $p_z \gg p_\perp$, a region which does not
receive large occupancy in the initial conditions.
Instead, these plasma-unstable modes presumably
start with typical occupancy $f \sim \half$, the amount provided by
vacuum fluctuations.%
\footnote{%
    Some readers might worry that it is cheating to seed a classical
    field with initial conditions set by the size of quantum
    fluctuations.  For a field which experiences an exponential
    instability, it is not cheating, a point discussed at some length in the
    inflationary preheating literature, see for instance Ref.\
    \cite{Khlebnikov}
    }
The plasma-unstable fields have little impact on the dynamics until they
have grown to occupancy
$\sim \alpha^{-1}$.  This requires a time of order
$t \sim \Qs^{-1} \ln^2(1/\alpha)$, or
$a \sim 2 \ln(\ln(1/\alpha))/\ln(1/\alpha)$.  Therefore, there is a
short delay before the plasma instabilities start to do their job.
During this time, Bjorken
expansion raises $d$, moving it away from the $d=c-1$ line.  So when the
system starts to become more isotropic, it does so with a mean occupancy
below $c=1$, arriving away from the $c=1,d=0$ point.  We illustrate this
idea in Figure \ref{fig_logs}.  By the time $d$ approaches 0, the two
attractor lines are separated, and we naturally land from ``above''
(the anisotropic side) on attractor 1.

Of course, it might somehow occur that the amplitude of plasma
instabilities overshoots the level where nonlinear interactions normally
limit instability growth.  That would allow the unstable fields to grow
larger, and angle change to be faster, than the estimates given above
(which are based on steady-state plasma instability evolution
\cite{KM1}).  For instance, it is known that, if the system is very
anisotropic and
if instabilities start with extremely small initial amplitudes, this can
occur \cite{RRS,BodekerRummukainen}.  But we claim that even such an
``overshoot'' cannot lead to near-isotropy.

First, note that the plasma-unstable fields have much larger magnetic
than electric field amplitude.  Therefore, they deflect hard excitations
more efficiently and they absorb their energy.  The time scale for
large-angle deflection is shorter than the time scale for large energy
absorption; so the plasma-unstable fields never dominate the energy
density, which remains in the hard excitations.  So the only way to get
isotropy is to get large-angle deflection.

We showed above that the scales $m$ and $\Qs$ become
separated; $m \ll \Qs$ (at least by $\log(1/\alpha)$).
The unstable fields have wave numbers $k \sim m$.
For a field of wave number $ k \sim m \ll \Qs$ to bend
excitations of momentum $\Qs$ by and order-1 angle in a time scale $1/m$,
the force would have to be $\dd p/\dd t \sim m \Qs$, requiring a magnetic
field of strength $B \sim m\Qs/g$.  But in nonabelian field theory,
whenever a magnetic field of wave number $k$ exceeds the field strength
$B \geq k^2/g$, it becomes Nielsen-Olesen unstable
\cite{NO}, leading to the very rapid ($t \lsim 1/k=1/m$) collapse of any
such magnetic field.  Therefore the magnetic fields associated with
plasma instabilities can never grow larger than $B \sim m^2/g$ -- too
small to deflect hard $p\sim \Qs$ excitations by large angles.

In the remainder of the paper we will concentrate on filling in all
details about the development of the system, assuming it follows the
first attractor.

\section{Detailed evolution along the first attractor}

\begin{figure}
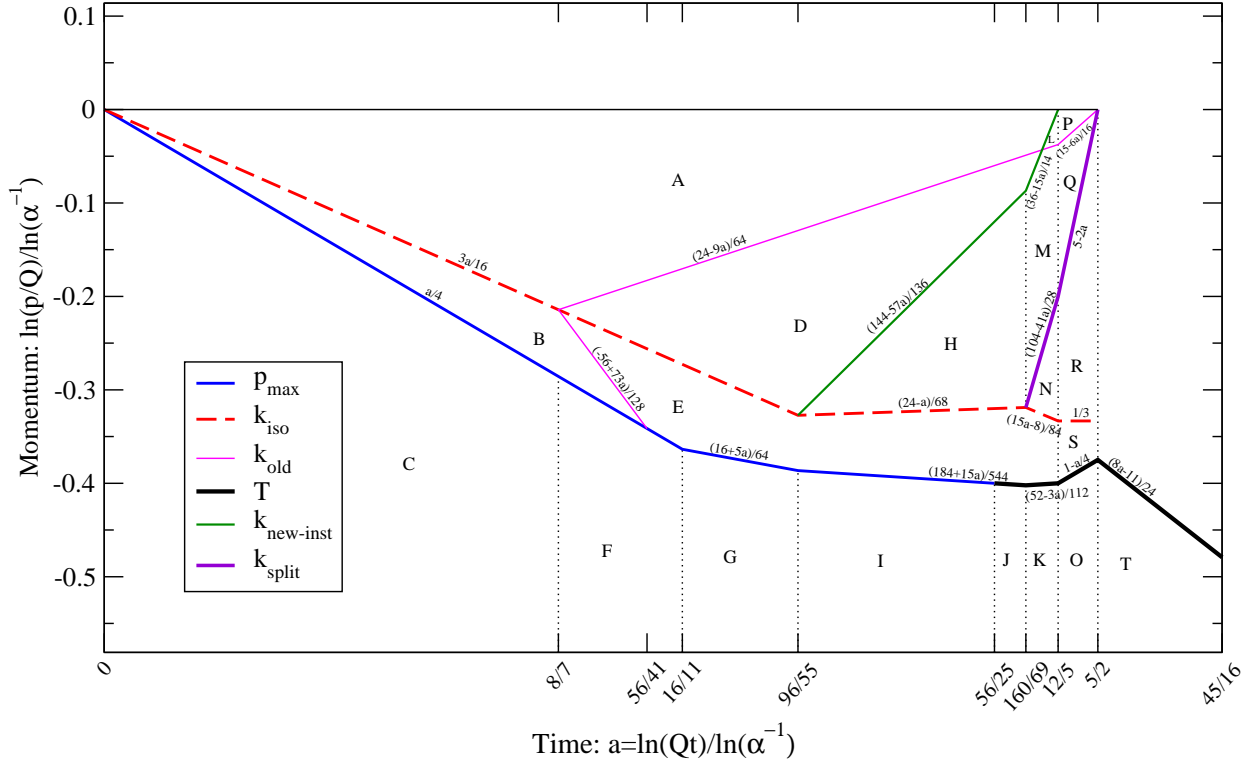

\centerbox{1.0}{fthetav2.eps}
\caption[Angles and occupancies]
{\label{fig1}
  Log-log plot of the time-momentum plane, divided into regions with
 distinct behaviors for the angular range of excitations about the $xy$
 plane, $\theta$, and the occupancy $f(p)$.  The horizontal axis
 is $a$ defined as $t = \alpha^{-a} \Qs^{-1}$ or $a=-\ln(t\Qs)/\ln(\alpha)$.
 The vertical axis is the log of momentum.  The equations describing
 lines are written in shorthand, so for instance $(8-3a)/16$ means
 $p = \Qs \alpha^{(8-3a)/16}$.  The letters $A$--$T$ refer to Table
 \ref{table1}.}
\end{figure}

In the previous Section, we described how the occupancies $c$ and the anisotropy
parameter $d$ of the primary particles with $p\sim \Qs$ evolve with time along the 
attractor 1, until at the time $a=12/5$, when the momentum diffusion experienced
by the primary particles is taken over by the soft sector. 
In this Section we describe the details how the soft sector evolves
during this time, and we describe in detail what happens after the soft
sector takes over the dynamics.
In particular, we find the occupancies and angular ranges of all excitations throughout
the evolution of the system. The results of this Section are collected in Figure \ref{fig1}
and Table \ref{table1}.
 
We divide the evolution in three main stages. 
In the early stage the primary particles have occupancies larger than 1, 
and all processes involving  these modes are stimulated. 
During this stage, the system grows steadily 
more anisotropic and radiated daughters become less and less important.  Incidentally, this means that 
classical Yang-Mills theory under Bjorken expansion never isotropizes and 
the interactions become weaker with time.

At the time $a=8/7$, the attractor 1 crosses the $c$-axis in Figure \ref{attractors}, and the 
occupancies of the hard excitations become small.  Therefore, the Bose stimulation 
of processes involving hard excitations is lost.  This is the middle stage. 
The loss of stimulation makes the interaction rates decrease \emph{more
  slowly}, and the soft sector will start to affect the dynamics.
The soft sector is also anisotropic and therefore has
its own unstable modes.  During the second
stage, the momentum diffusion%
\footnote{%
    Here and in the following, ``transverse'' or $p_\perp$ refers to the
    directions transverse to the initial momentum of the
    particle, not the directions transverse to the beam axis.}
$\qhat \equiv \dd p_\perp^2/\dd t$ arising from
these new plasma instabilities increases relative
to the ``primary'' instabilities of the hard particles.
The new plasma instabilities come to dominate $\qhat$ at large angles
at the time $a=96/55$.  But $\qhat$ at narrow angles, experienced 
by the primary excitations, remains dominated by the ``primary''
instabilities.
Most of the particle number in the soft sector is in a nearly isotropic bath, whose
distribution forms an $f(p) \propto 1/p$ ``tail'' below a
cutoff scale $\pmax$.  Initially $f(\pmax) \gg 1$, but
at the time $a=56/25$, $f(\pmax)=1$ and the soft bath becomes nearly
thermal.  

Plasma instabilities from the nearly-thermal bath grow to dominate at all
angles at time $a=12/5$, which is where attractor 1 in Figure
\ref{attractors} intersects with the boundary between
the regions labeled ``Plasma Instabilities'' and ``Soft Particle Bath
Forms''.  Thereafter, the broadening of the hard primary particles will
be controlled by the soft sector and its plasma instabilities.
The instabilities cause hard particles to undergo 
successive splitting processes, depositing their energy into the thermal
bath, increasing its temperature. At time $a=5/2$, the primary
$p\sim \Qs$ particles have had time to break up completely, leaving only
a nearly isotropic thermal bath.  We now discuss these stages in more
detail.

\begin{table}
\centerline{
 \begin{tabular}{|c|c|c||c|c|c|} \hline
 $\;\;$Region$\;\;$ & $\;\;\theta$ range$\;\;$ & 
 $\;\;$ Occupancy $f(p)\;\;$ &
 $\;\;$Region$\;\;$ & $\;\;\theta$ range$\;\;$ & 
 $\;\;$ Occupancy $f(p)\;\;$ \\ \hline
 A  & $\alpha^{\frac a8} (\Qs/p)^{\frac 23}$ & $\alpha^{\frac{-8+7a}{8}} (\Qs/p)^{\frac{1}{6}}$ & 
 K  & 1 &  $\alpha^{\frac{52-3a}{112}} (\Qs/p)^{1}$ \\
 B  & 1 & $\alpha^{\frac{-14+27a}{14}} (\Qs/p)^{\frac{81}{14}}$ &
 L  & $\alpha^{\frac{48-13a}{56}} (\Qs/p)^1$ & $\alpha^{\frac{-104+69a}{56}} (\Qs/p)^{\frac{-1}{6}}$ \\
 C  &  $1$  &  $\alpha^{\frac{-8+7a}{8}} (\Qs/p)^1$ &
 M  & $\alpha^{\frac{48-13a}{56}} (\Qs/p)^1$ &  $\alpha^{\frac{-6+6a}{7}} (\Qs/p)^{\frac{5}{2}}$ \\
 D  & $\alpha^{\frac a8} (\Qs/p)^{\frac 23}$ & $\alpha^{\frac{a}{2}} (\Qs/p)^{\frac{17}{6}}$ &
 N  & $\alpha^{\frac{-8+15a}{112}} (\Qs/p)^{\frac{3}{4}}$ & $\alpha^{\frac{8+55a}{112}} (\Qs/p)^{\frac{11}{4}}$ \\
 E  & 1 &  $\alpha^{\frac{5a}{8}} (\Qs/p)^{\frac{7}{2}}$ &
 O  & 1 &  $\alpha^{\frac{4-a}{4}} (\Qs/p)^{1}$ \\
 F  & 1 &  $(\Qs/p)^{1}$ &
 P  & $\alpha^{\frac{3-a}{2}} (\Qs/p)^1$ & $\alpha^{\frac{-5+3a}{2}} (\Qs/p)^{\frac{-1}{6}}$ \\
 G  & 1 &  $\alpha^{\frac{-80+55a}{128}} (\Qs/p)^{1}$ &
 Q  & $\alpha^{\frac{3-a}{2}} (\Qs/p)^1$ & $\alpha^{\frac{a}{2}} (\Qs/p)^{\frac{5}{2}}$ \\
 H  & $\alpha^{\frac{24-a}{68}} (\Qs/p)^{1}$ & $\alpha^{\frac{-48+87a}{136}} (\Qs/p)^{\frac{5}{2}}$ &
 R  & $\alpha^{\frac{1}{4}} (\Qs/p)^{\frac{3}{4}}$ & $\alpha^{\frac{5}{4}} (\Qs/p)^{\frac{11}{4}}$ \\
 I  & 1 &  $\alpha^{\frac{-920+605a}{1088}} (\Qs/p)^{1}$ &
 S  & 1 &  $\alpha^{\frac{3}{2}} (\Qs/p)^{\frac{7}{2}}$ \\
 J  & 1 &  $\alpha^{\frac{184+15a}{544}} (\Qs/p)^{1}$ &
 T  & 1 &  $\alpha^{\frac{8a-11}{24}} (\Qs/p)^{1}$ \\
 \hline
 \end{tabular}}
\caption[Results table]
 {\label{table1}
  Angular range of excitations and typical occupancies for the regions
  labeled in Figure \ref{fig1}.}
\end{table}

\subsection{Early stage:  $a<8/7$}

\subsubsection{Angular distribution, the scale $\kiso$}

We will now find the angular ranges and occupancies
of the soft sector during the evolution.
Plasma instabilities give rise to angle dependent transverse momentum diffusion
characterized by $\qhat$ of size \cite{KM1}
\be
\label{qhat1}
\qhat \sim \left\{ \begin{array}{ll}
  \delta^{-2} m^3 \sim \alpha^{\frac{5a}{4}} \Qs^3 \,, &
    p_z \lsim \delta p_\perp\,, \; 
    p>\delta^{-2} m \sim \alpha^{\frac{a}{4}} \Qs \,, \\
  \delta^{-1} \theta^{-1} m^3 
  \sim \theta^{-1} \alpha^{\frac{11a}{8}} \Qs^3 \,, &
    p_z/p_\perp \equiv \theta \,, \;
    p> m/(\delta \theta) \,, \\
 \end{array} \right.
\ee
where $m$ is the thermal mass
\be
\label{msq1}
m^2 \sim \alpha \int \frac{\dd^3 p}{p}f(p) \sim \alpha^{1-c+d} \Qs^2 \sim \alpha^{a} \Qs^2 \,, \qquad
m \sim \alpha^{a/2} \Qs \,.
\ee

The angular range which excitations of momentum
$p$ occupy will be given by
\be
\theta^2 p^2 \sim \qhat(\theta) t 
  \sim \alpha^{\frac{11a}{8}} \theta^{-1} \alpha^{-a} \Qs^2
\qquad \Rightarrow \qquad
\theta \sim \alpha^{\frac{a}{8}} (\Qs/p)^{\frac{2}{3}} \,.
\label{eq:theta}
\ee
We define the scale where $\theta \sim 1$ as $\kiso$:
\be
\label{kiso1}
1 \sim \alpha^{\frac{a}{8}} (\Qs/\kiso)^{\frac{2}{3}}
\qquad \Rightarrow \qquad
\kiso   \sim \alpha^{\frac{3a}{16}} \Qs \,.
\ee
Below this scale particles are nearly isotropic.

\subsubsection{Particle production}
The instabilities not only
change the angular range the soft particles populate; they also affect
their occupancies through instability induced splitting processes. 
Provided $p \geq \delta^{-2} m \sim \alpha^{\frac{a}{4}} \Qs$, particle
emission is LPM suppressed.  The formation time for an emission is
\be
\label{tform1}
\tform^{-2}(p,\theta) \sim \qhat(\theta) p^{-1} 
  \sim \alpha^{\frac {5a}{4}} \Qs^3 p^{-1} (\delta/\theta)
\ee
but for our case, in a time $\tform$ the radiated daughter only
accumulates a transverse
momentum of $\sqrt{\qhat \tform} < \delta p$, so
$\theta \sim \delta$.  Ignoring stimulation, the rate each hard particle
splits off daughters of momentum $p$ is
$d\Gamma/d\ln(p) \equiv \Gsplit \sim \alpha \tform^{-1}$.  Summing over
the number density hard particles $\nhard$ and including the final-state
hard particle stimulation factor $[1{+}f(\Qs)]\sim \alpha^{-c}$,
the number of daughters radiated per logarithmic momentum range in
time $t\sim \alpha^{-a} \Qs^{-1}$ is
\be
\label{n1}
n_{\rm daughter}(p,a) \sim \nhard \alpha^{-c} \alpha \tform^{-1} t
     \sim \alpha^{d-c} \Qs^3 \alpha^{1-c} \alpha^{\frac{5a}{8}} 
          \alpha^{-a} (\Qs/p)^{\frac 12}
     \sim \alpha^{-1+\frac{3a}{2}} \Qs^{\frac 72} p^{-\frac 12} \,.
\ee

\subsubsection{Redshifting and old particles}

As a consequence of expansion, the rate of particle
production falls quickly ($\propto \alpha^{\frac{3}{2}a}$), and 
therefore the distribution of soft particles at a given time $a$ need
not to be dominated by the particles emitted during that time scale.
The daughters radiated at an earlier time $a'$ are still around, though
they have reduced in number and energy due to Bjorken expansion.

For $p>\kiso$, 
the longitudinal momentum component is parametrically small
compared to the total momentum, and the
effect of the expansion 
is simply to reduce the number of particles by a factor $\alpha^{a-a'}$.
So the number density of particles, emitted at time $a'$, which are
present at time $a>a'$ is
\be
n_{\rm redshift}(p>\kiso,a;a') \sim \alpha^{a-a'}n_{\rm daughter}(p,a').
\ee
For $p<\kiso$, the expansion also decreases the $z$-component 
of momentum. The plasma instabilities keep the soft distribution
below $\kiso$ isotropic so that the net effect of the expansion and
angle randomization is to redshift the isotropic spectrum below $\kiso$ to 
smaller momenta, much like in the case of attractor 2;
 \be
n_{\rm redshift}(p<\kiso,a;a')\sim  \alpha^{a-a'} 
  n_{\rm daughter}(\alpha^{-\frac{1}{3}(a-a')} p,a')\,.
 \ee

Both above and below $\kiso$, the particle number density at momentum scale $p$
is dominated by particles which where emitted at the scale $\kiso$. 
For $p>\kiso$,  the dominant contribution to $n(p)$ arises from time $a'$ when
$\kiso$ was $p$, namely $\alpha^{a'}\sim (p/\Qs)^{\frac{16}{3}}$, so that the 
particle number density above $\kiso$ is
\be
n(p) 
\sim\alpha^{a-a'}\alpha^{-1+\frac{3}{2}a'}\Qs^{\frac{7}{2}}p^{-\frac{1}{2}}
\sim \alpha^{-1+a}(p/\Qs)^{\frac{13}{6}}\Qs^3,
\label{for_C}
\ee
and the corresponding occupancies in the occupied angle region are 
\be
f(p)\sim \frac{n(p)}{\theta(p) p^3}\sim 
  \alpha^{-1+\frac{7}{8}a}(\Qs/p)^{\frac 16}.
\ee
This is Region A in Figure \ref{fig1}.

For $p<\kiso$, the particles created at the scale $\kiso$ at time $a'$ 
have just redshifted to scale $p$ at time $a$ provided that
\be
\alpha^{\frac{1}{3}(a-a')} \kiso(a') \sim p,
\ee
so that the particle number density at time $a$ is dominated by 
particles created at the time
\be
\alpha^{a'}\sim \alpha^{\frac{16}{7}a}(\Qs/p)^\frac{48}{7}.
\ee
The particle number density and occupancies then read
\begin{eqnarray}
\label{for_B}
n(p)
&\sim& \alpha^{a-a'}\alpha^{-1+\frac{3}{2}a'}\Qs^{\frac{7}{2}}(\alpha^{-\frac{1}{3}(a-a')}p)^{-\frac{1}{2}}
\sim \alpha^{-1+\frac{27}{14}a}(\Qs/p)^{\frac{39}{14}}\Qs^{3},\\
f(p)&\sim& \frac{n(p)}{p^3} \sim \alpha^{-1+\frac{27}{14}a}(\Qs/p)^{\frac{81}{14}}.
\end{eqnarray}
This is Region B in Figure \ref{fig1}.

\subsubsection{Saturation, the scale $\pmax$}
As seen in KM1, distributions that have 
infrared tails steeper than that of a thermal distribution
rearrange themselves very effectively to form a thermal-like tail so that
\be
f(p)\propto 1/p, \quad \textrm{ for } p<\pmax.
\ee
 In particular, if the 
occupancies of daughters $f(p)$ exceed
\be
f(p)\sim \frac{\Qs}{p}[1{ + }f(\Qs)], \label{eq:fmax}
\ee
the rate for the daughter particles to \emph{rejoin} the hard particles
becomes larger than the corresponding emission rate, and the soft particle number
density in fact \emph{decreases} due to inelastic scatterings. This is the case in the 
Region B below $p\sim \alpha^{\frac{59a}{268}}\Qs$. The total rate for soft
particles rejoining to hard particles is (see Eq.~(2.54) of KM1)
\be
\label{Grejoin}
\Gamma_{\rm rejoin}(p)\sim \frac{1}{n(p)}\int_{k\sim \Qs} 
    \dd^3k \; \Gamma_{\rm split}(p)\frac{p}{k}f(k)f(p)
\;\sim\; \frac{\nhard}{\Qs}\frac{ \Gamma_{\rm split}(p)}{p^2} 
\sim \alpha^{\frac{8+13a}{8}}\Qs (\Qs / p)^{\frac 52},
\ee
so that the modes at scales below $\pmax$, with 
$\Gamma_{\rm rejoin}(\pmax) t \lesssim 1$  or
\be
\label{pmax1}
\pmax \sim \alpha^{\frac{a}{4}}\Qs
\ee
have had time to 
change their occupancies by an order 1 amount and adjust to
the form of Eq.~(\ref{eq:fmax}). Below $\pmax$ the 
occupancies then read
\be
f(p)\sim \alpha^{-1+\frac{7}{8}a}(\Qs/p).
\ee 
This is the region C of Figure \ref{fig1}.

\subsection{Middle stage: $8/7<a<12/5$}

\subsubsection{$a > 8/7$ New particles, the scale $\kold$}
For $a>8/7$, the Bose stimulation for the particle production is lost.
As a result, the particle production rate decreases more slowly and particles
produced at later times may start to dominate the particle number density.
The number of soft particles created at the time scale $a$ is now
(remove $\alpha^{-c}$ from \Eq{n1})
\be
\label{new_daughter}
n_{\rm daughter}(p,a)\sim \nhard \alpha \tform^{-1} t
\sim\alpha^{\frac{5a}{8}}\Qs^3(\Qs/p)^{\frac{1}{2}}.
\ee
For $k>\kiso$ this becomes dominant over the redshifted 
particles for $k<\kold$ given by
(compare \Eq{new_daughter} with \Eq{for_C})
\be
\kold \sim \alpha^{\frac{3}{8}-\frac{9a}{64}}\Qs.
\ee
For $k<\kiso$ the new particles become more numerous for $k>\kold$ with
(\Eq{new_daughter} and \Eq{for_B})
\be
\kold \sim \alpha^{-\frac{7}{16}+\frac{73 a}{128}}\Qs.
\ee 
These are the regions D and E in Figure \ref{fig1}, respectively.
Region B ends at $a=\frac{56}{41}$.

\subsubsection{$a > 16/11$: Soft joining, the scale $\pmax$}
\label{seq:softjoining}
We defined the scale $\pmax$ as the one below which 
$f(p<\pmax) \propto 1/p$.  At early times we saw $\pmax$ is set by the
rate of emission/absorption off hard $p\sim \Qs$ excitations.  At later
times, merging processes between $p\sim \pmax$ excitations can become
faster and come to control this scale.
The rate for soft joining (in Region E, just above $\pmax$) is
\be
\Gmerge(p) \sim \alpha \tform^{-1}(p,\theta \sim 1) [1{+}f(p)]
\sim \alpha^{1+\frac{21 a}{16}}(\Qs/p)^4\Qs \,,
\ee
and $\Gmerge(\pmax) t > 1$, using $\pmax$ from \Eq{pmax1}, after
time $a = 16/11$.
After this, the evolution of $\pmax$ is controlled by soft merging;
\be
\Gmerge(\pmax) t\sim 1 \qquad \Rightarrow \qquad  \pmax \sim
\alpha^{\frac{1}{4}+\frac{5a}{64}}\Qs \,.
\ee

Modes with $p\lsim \pmax$ effectively thermalize with each other.  The
energy density is dominated by $p\sim \pmax$ modes, so the occupancy is
determined by the energy density of particles residing at the scale
$\pmax$.  Hence, for $p<\pmax$,
\be
f(p)\sim \frac{f(\pmax)\pmax}{p}\sim \alpha^{\frac{-5}{8}+\frac{55a}{128}}(\Qs/p).
\ee
This is Region G in Figure \ref{fig1}.

\subsubsection{$a > 96/55$: New plasma instabilities, the scale  $\knewinst$}

The distribution of daughter particles is anisotropic; therefore the
daughter particles also give rise to plasma instabilities. As discussed in
KM1, the strongest instabilities are driven by daughters at the scale
$\kiso$, where the angular distribution is order-1 anisotropic.  The
resulting plasma instabilities give rise to momentum diffusion
\be
\qhatnew \sim m^3(\kiso) \sim \alpha^{\frac{3}{2}} f^{\frac{3}{2}}(\kiso) \kiso^3,
\label{qhatnew}
\ee
which is comparable in all angular directions $\theta$ (see
Eq.~(4.21-4.22) of KM1).  By comparing \Eq{qhat1} to \Eq{qhatnew},
using \Eq{kiso1} and \Eq{new_daughter}, we find that
$\qhatnew = \qhatinst$ at large angles $\theta \sim 1$ at $a=96/55$.
After this, the evolution of the scale $\kiso$ is driven by the instabilities
caused by the scale $\kiso$ itself.
The scale $\kiso$ can be found by self-consistently solving
\be
\kiso^2 \sim \qhatnew t \quad \mbox{and} \quad
\qhatnew \sim \alpha^{\frac 32} f^{\frac 32}(\kiso) \kiso^3
\quad \mbox{with} \quad
f(\kiso) \sim \alpha^{\frac{5a}{8}} \Qs^{\frac 72} \kiso^{-\frac 72}
\ee
resulting in
\be
\kiso \sim \alpha^{\frac{24-a}{68}} \Qs \,, \qquad
\qhatnew \sim \alpha^{\frac{24+33a}{34}} \Qs^3 \,, \qquad
f(\kiso) \sim \alpha^{\frac{-42+23a}{34}} \,.
\ee

When $\qhatnew$ dominates, particles at scales above $\kiso$ will have
an angular range
$
\theta \sim \alpha^{\frac{24-a}{68}}(\Qs/p) \,.
$
The range of angles where $\qhatnew > \qhatinst$ is
$
\theta > \alpha^{\frac{55a-96}{136}} \,.
$
Using \Eq{eq:theta}, we see that excitations experience $\qhatnew$
provided $p < \knewinst \sim  \alpha^{\frac{144-57a}{136}} \Qs$. 
This is the region H in Figure \ref{fig1}.

The new $\qhat$ changes the rate of the soft merging and therefore the 
functional form of $\pmax$ as well as $f(p<\pmax)$. Repeating
the analysis of Subsection \ref{seq:softjoining} with $\qhat\sim \qhatnew$ gives 
\be
\pmax \sim \alpha^{\frac{184+15a}{544}} \Qs
\quad \textrm{ and } \quad
f(p)\sim \alpha^{\frac{605a-920}{1088}}(\Qs/p).
\ee
This is the region I in Figure \ref{fig1}.

\subsubsection{$a > 56/25$: Thermalization of the soft sector and the scale $\ksplit$}

At the time $a= 56/25$, $f(\pmax)$ drops to $\sim 1$ and
the cascade of the infrared modes by joining processes, which started
at the time scale $a=16/11$, is complete. The soft sector now forms a
nearly-thermal bath with a temperature $T\sim \pmax$.
Subsequently, the dominant inelastic process changes from joining
to splitting, and those particles with $p>T$ that have had time to
undergo a democratic splitting (\ie, emit a daughter with a momentum
comparable to emitter's momentum)
in less than the system age cascade down in energy by multiple splittings, 
and lose their energy to the thermal bath.  
The modes which have had time to undergo a democratic splitting are
those for which $p<\ksplit$, with
\be
\Gsplit(\ksplit) t \sim  \alpha \tform^{-1} t \sim 1 
 \quad \Rightarrow  \quad 
\ksplit \sim \alpha^2 \qhatnew t^2 \sim \alpha^{\frac{92-35a}{34}}\Qs\,.
\ee
The temperature of the bath is set by the highest momentum modes falling
onto it, so that the temperature becomes
\be
T \sim \varepsilon^{\frac{1}{4}} \sim
(f(\ksplit)\ksplit^4)^{\frac{1}{4}}\sim \alpha^{\frac{184+15a}{544}}\Qs \,.
\ee
This explains Region J in Figure \ref{fig1}.

Most of the excitations below $\kiso$ but above $T$ have had time to split
and release their energy to the thermal bath. However, a small portion
of the particles, the ones created in the last time scale $t_{\rm residence}\sim\Gsplit^{-1}(p)\sim(p/\ksplit)^{1/2}t$, are still
around and have not yet undergone the cascade to lower scales. 
The flux of energy from 
the scale $\ksplit$ moving through scales to the scale $T$ is 
\be
\frac{\dd\varepsilon}{\dd t}\sim \varepsilon(\ksplit)/t \sim f(\ksplit) \ksplit^4/t,
\ee
so that the energy density carried by particles created in the last
$t_{\rm residence}$ time scale at the momentum scale $p$ is
\be
\varepsilon(p)\sim t_{\rm residence}\frac{\dd \varepsilon}{\dd t},
\ee
so that $f(p)$ and $n(p)$ for $\kiso > p > T$ are
\be
f(p) \sim \varepsilon(p)/p^4 
     \sim f(\ksplit)(\ksplit/p)^{\frac 72}
     \sim \alpha^{\frac{5a}{8}}(\Qs/p)^{\frac{7}{2}}, 
     \qquad n(p)\sim \alpha^{\frac{5a}{8}}(\Qs/p)^{\frac{1}{2}}\Qs^3
\label{eq:fDprime}
\ee
which is a continuation of Region E.


\subsubsection{$a > 160/69$: Instabilities from the nearly thermal bath}

\label{seq:instab}

At time $a=160/69$ the scale $\ksplit$ crosses the scale $\kiso$.
This reduces the number of excitations at the scale $\kiso$, which
ceases to dominate as a source of plasma instabilities.  However, the
thermal bath is incompletely isotropic, so it also gives rise to plasma
instabilities, with associated momentum broadening parameter
$\qhatTinst$.  Beyond $a=160/69$, $\qhatTinst$ dominates for 
$\theta\sim 1$.

Two mechanisms keep the
soft nearly-thermal bath incompletely isotropic.  First, Bjorken expansion
continually ``stretches'' the thermal bath. Secondly, because $\qhatnew$
is order-1 anisotropic, the radiated daughters which arrive at the scale
$T$ do so with an $\OO(1)$ small-$\theta$ bias.
Both mechanisms return a level of anisotropy for the bath of
\be
\label{epsilon1}
\epsilonbath \sim \frac{T^2}{\qhatTinst t} \,,
\ee
which is the ratio of a typical $p_\perp^2$ in the bath to
the change in $p_\perp^2$ during the system's age.

Using Eq.~(3.16) of KM1, we find
\be
\qhatTinst \sim \epsilonbath^{\frac 32} m(T)^3 \sim
 \epsilonbath^{\frac 32} \alpha^{\frac 32} T^3 \,.
\ee
Substituting $\qhatTinst$ into \Eq{epsilon1}, we find
\be
\epsilonbath \sim \alpha^{-\frac{3}{5}} (tT)^{-\frac{2}{5}} \,, \qquad
\qhatTinst \sim \alpha^{\frac{3}{5}} T^{\frac{12}{5}} t^{-\frac{3}{5}} \,.
\label{qhatlate}
\ee
The temperature is still determined by the $\ksplit$ scale, 
$T \sim(f(\ksplit))^{1/4}\ksplit$, and the splitting scale is given by
$\ksplit \sim \alpha^2 \qhatTinst t^2$.
Solving self-consistently for $\epsilon$, $\qhatTinst$, $\ksplit$, and $T$
gives
\be
T \sim \alpha^{\frac{13}{28} - \frac{3a}{112}} \Qs \,, \qquad
\qhatTinst \sim \alpha^{\frac{12}{7} + \frac{15a}{28}} \Qs^3 \,, \qquad
\ksplit \sim \alpha^{\frac{26}{7}-\frac{41a}{28}} \Qs \,,  \qquad
\epsilonbath \sim \alpha^{\frac{-44+23a}{56}} \,.
\ee
This explains Region K in Figure \ref{fig1}.

The new dominant $\qhat$ changes the rate of angular broadening
and hence also the scale $\knewinst$.
The momentum range where $\qhatnew$ dominates, and the angular range of
excitations, are
\be
\qhatTinst > \qhatinst \mbox{ if } p< \knewinst \sim \alpha^{\frac{36-15a}{14}} \Qs 
\quad \mbox{where} \quad
\theta \sim \alpha^{\frac{48-13a}{56}} \Qs/p \,.
\ee
This explains Regions L and M, above and below $\kold$, respectively. 

Below $\ksplit$ but above $T$ the cascading particles start their 
fall in a small angle range at the scale $\ksplit$ and as they cascade down
in energy, their angular range broadens. The cascading particles 
stay at the momentum scale $p$ for the time $t_{\rm residence}\sim (p/\ksplit)^{1/2}t$,
and during this time their angular range grows to 
\be
p^2 \theta^2 \sim \qhatTinst t_{\rm residence}  \qquad \Rightarrow \qquad
\theta \sim \alpha^{\frac{-8+15a}{112}} (\Qs/p)^{\frac{3}{4}}
\ee
with $\theta \sim 1$ for $p \lsim \kiso \sim \alpha^{\frac{-8+15a}{84}} \Qs$.
The estimate for the particle number density from Eq.~(\ref{eq:fDprime})
still holds so that the occupation number below $\ksplit$ is
$
f(p)\sim \alpha^{\frac{8+55a}{112}}(\Qs/p)^{\frac{11}{4}}.
$
This is Region N in Figure \ref{fig1}.%
\footnote{A quick calculation of $\qhat$ due to instabilities associated
  with the scale $\kiso$ returns
  $\qhatnew \sim \alpha^{\frac{48+15a}{28}} \Qs^3 \sim \qhatTinst$.
  All scales from $\kiso$ down to $T$ actually have comparable roles in
  setting $\qhat$; but this does not change parametric estimates.}

\subsection{Late stages: $12/5<a<5/2$}

At $a=12/5$ we find $\qhatinst \sim \qhatTinst$.  Beyond this
scale, the new plasma instabilities from the incompletely isotropic
thermal bath dominate momentum broadening at {\sl all} scales, \ie, this 
is the first moment when the soft sector starts to influence the dynamics of the hard
particles. This is the point $d=3/10$, $c=-11/10$ in Figure
\ref{attractors}, where attractor 1 enters the
region marked ``Soft Particle Bath Forms.''
As a result, the anisotropy of the hard modes departs from $d=a/8$ to become 
more isotropic, and the rate of  radiation of daughters increases, since
it is now controlled by the larger $\qhatTinst$.

The splitting scale is now given by $\ksplit \sim \alpha^{2} \qhatTinst t^{2}$.
Since there is only a single $\qhat$, $\ksplit$ is the scale where there
is of order 1 daughter per hard particle.  So
\be
T^4 \sim \nhard \ksplit \sim \alpha^{1-a} \qhatTinst \Qs
\ee
while \Eq{qhatlate} still holds.
Solving solving self-consistently, we find
\be
\qhatTinst \sim \alpha^3 \Qs^3 \,, \qquad
T \sim \alpha^{1-\frac{a}{4}} \Qs \,, \qquad
\ksplit \sim \alpha^{5-2a} \Qs \,, \qquad
\epsilonbath \sim \alpha^{\frac{a-2}{2}} \,.
\ee
This is Region O in Figure \ref{fig1}.

The new $\qhat$ broadens the angular distribution more effectively,
and for $p>\ksplit$ we find
$\theta(p) \sim \alpha^{\frac{3-a}{2}} \Qs/p$. In particular, the angular 
range of the primary hard excitations is 
$\delta \sim \alpha^{\frac{3-a}{2}}$,
which grows with time for the first time since $a\simeq 0$.
However, the hard excitations remain anisotropic to the bitter end.

Repeating the calculation of the behavior of particles within the
cascade, we get the results for regions P, Q, R, and S in Figure
\ref{fig1}.

The thermal bath continues to draw energy from the 
hard excitations until $\ksplit$ reaches $\Qs$ and the hard excitations,
which dominate the system's energy, themselves split and join the
thermal bath. This occurs at $a=5/2$.  At this stage the temperature is
$T \sim \alpha^{\frac{3}{8}} \Qs$, and the residual anisotropy of the
nearly-thermal bath is $\epsilon \sim \alpha^{\frac{1}{4}}$.  This is still
larger than the value $\epsilon \sim \alpha^{\frac{1}{3}}$ where a thermal bath
becomes dominated by scattering rather than plasma instabilities.

A shortcut calculation which leads to the correct final equilibration
time is the following.  The energy density scales with time as
$\varepsilon \sim \alpha^{a-1} \Qs^4$, so the would-be equilibration
temperature is $T(a) \sim \alpha^{\frac{a-1}{4}} \Qs$.  According to KM1
\cite{KM1}, the equilibration time for an anisotropic system with
$T<\Qs$ is
\be
\teq \sim \alpha^{\frac{-13}{7}} \Qs^{\frac 57} T^{\frac{-12}{7}} \,.
\ee
(This result was derived by asking how long it takes a thermal bath at
temperature $T$ to cause hard splitting in excitations of momentum $\Qs$,
given that the thermal bath will be somewhat anisotropic and the
dynamics are dominated by plasma instabilities; see KM1 \cite{KM1}.)
Equating $\teq$ with the age of the system $t \sim \alpha^{-a} \Qs^{-1}$
and inserting our expression for $T$, we find $a=5/2$ or
$\teq \sim \alpha^{\frac{-5}{2}} \Qs^{-1}$.

\subsection{Aftermath:  $a>5/2$}

After $a=5/2$ the $p\sim \Qs$ initial excitations are gone, leaving behind
a nearly thermal bath. The system is again characterized by a single scale, 
the temperature $T$, and hence the system 
can again be described using Figure \ref{attractors}. At the time $a=5/2$,
$d=-1/4$ and $c=0$; subsequent evolution makes the distribution 
more isotropic, and the solution will move down along the $d$-axis.

The system is now nearly isotropic, and therefore
the energy density begins to scale with time as 
$\varepsilon \propto t^{-4/3}$ rather than as $t^{-1}$.  Using the value
of the energy density at $a=5/2$, we find
\be
T \sim \alpha^{\frac{a}{3} - \frac{11}{24}} \Qs \,.
\ee
Initially $\qhat$ remains dominated by plasma instabilities from the
residual anisotropy of the thermal bath giving rise to ``anomalous
viscosity''.  We
still find \Eq{qhatlate} to hold; substituting the known $T$ value,
\be
\qhat \sim \alpha^{\frac{7a}{5} - \half} \Qs^3 \,,
\qquad
\epsilon \sim \alpha^{\frac{4a}{15} - \frac{5}{12}} \,.
\ee
This drops below the value $\epsilon \sim \alpha^{1/3}$, entering the regime
where ordinary scattering dominates, at
$a=45/16$.  Beyond this time scale, the system behaves like
an ordinary thermal bath expanding with a viscosity given by the normal
(non-anomalous) viscous law; 
$\qhat \sim \alpha^2 T^3 \sim \alpha^{a+\frac{5}{8}} \Qs^3$ and the
residual anisotropy is
$\epsilon \sim \alpha^{\frac{2a}{3} - \frac{37}{24}}$.

In comparison, attractor 2 reaches the $d$-axis
at the time $a=135/56$, so that the system becomes nearly thermal 
slightly earlier than in the case of the first attractor. 
As a consequence of being nearly isotropic from the beginning, 
the temperature is now only
\be
T_{\att2} \sim \alpha^{\frac{a}{3}-\frac{1}{4}}\Qs
\ee
resulting in
\be
\qhat_{\att2} \sim \alpha^{\frac{7a}{5}}\Qs^3\,,
\qquad
\epsilon_{\att2} \sim \alpha^{\frac{4a}{15}-\frac{1}{2}}.
\ee
In the case of attractor 2, scattering comes to
dominate at the time $a=25/8$, after which $\qhat_{\att2} \sim \alpha^{a+\frac{5}{4}}\Qs^3$ and 
$\epsilon_{\att2}\sim \alpha^{\frac{2a}{3}-\frac{7}{4}}$.

In both cases, the behavior after scattering comes to dominate 
continues, in principle, indefinitely.  At exponentially
late times, 
$a \simeq 3\alpha^{-1}/\beta_0 \ln(\alpha^{-1})$,
the temperature scale reaches
$\Lambda_{\rm QCD}$ and hadronization occurs.

\section{Discussion}

We have presented a complete parametric description of thermalization
after a heavy ion collision in the theoretically clean, in-principle
limit of arbitrarily large and high-energy ions, such that the scale
$\Qs$ (which dominates particle production from the original collision)
is taken so large that $\alphas(\Qs) \ll 1$.  Our most significant
finding is that thermalization does eventually occur.  Plasma
instabilities dominate the dynamics at all times until well after the
system becomes nearly isotropic and thermal.  Initially, they cause the
highly anisotropic starting configuration to become order-1 isotropic.
But Bjorken expansion, and falling typical occupancies, allow the system
to again become anisotropic -- though the level of anisotropy only grows
as $t^{\frac{1}{8}}$.

The process of thermalization is dominated by
\begin{enumerate}
\item
the generation of plasma instabilities due to the anisotropy of the
typical $p\sim \Qs$ excitations;
\item
the radiation of daughter excitations, which eventually form a
nearly-thermal bath;
\item
the plasma instabilities generated by the residual anisotropy of these
daughter excitations, which are responsible for finally breaking up the
$p\sim \Qs$ excitations and thermalizing the system at a time scale
$t \sim \alpha^{-\frac{5}{2}} \Qs^{-1}$.
\end{enumerate}
After near-thermalization the system remains dominated by plasma
instabilities of a nearly-thermal bath (anomalous viscosity) until the
time scale $t\sim \alpha^{-\frac{45}{16}} \Qs^{-1}$, after which elastic
scattering becomes the most important physical process and the plasma is
well described by standard kinetic theory.

Since there is an extended period of expansion in which the equation of
state is $\varepsilon \propto t^{-1}$ rather than
$\varepsilon \propto t^{-\frac{4}{3}}$ as for a thermal system, there is
significant entropy generation.
Most of the entropy of the final system is generated around
$t\sim \alpha^{-\frac{5}{2}} \Qs^{-1}$, when the initial hard excitations fragment
and join the thermal bath.  This entropy generation could have
phenomenological consequences in connecting the Colored Glass initial
conditions to final state multiplicities, see
\cite{BMSS2}.

It is not clear to us how high $\Qs$ would actually need to be for the
parametric estimates of this paper really to make sense.  However we
feel that this work does resolve the long-standing puzzle of what
happens in a heavy ion collision in what should be the most
theoretically clean limit -- that of extremely large saturation scale.

It would be nice to replace this {\sl parametric} estimate with a 
{\sl numerical} estimate, by computing the ``constant'' in
$t_{\rm therm} = (\#) \alpha^{\frac{-5}{2}} \Qs^{-1}$.  The good news is
that this only requires one to study the final stage of thermalization,
$12/5 < a < 5/2$ (since all other stages take
parametrically less time).  To study the relevant physics in this stage,
we would need to determine the relationship between anisotropy $\epsilon$
and momentum diffusion $\qhat$ for a nearly thermal bath, and we would
need to compute the exact rate of fragmentation for a hard excitation
experiencing a given value of $\qhat$.  The former could be done by
studying the plasma instabilities of a weakly anisotropic system.  This
should be doable using existing methods which combine classical field
dynamics with anisotropic hard loops.  The latter is also
straightforward.

\section*{Acknowledgements}

This work was supported in part
by the Natural Sciences and Engineering Research Council of Canada and
the Institute of Particle Physics (Canada).

\end{document}